# Applying Standards to Advance Upstream & Downstream Ethics in Large Language Models

J. Berengueres, *Member IEEE,* and M. Sandell

*Abstract*— This paper explores how AI-owners can develop safeguards for AI-generated content by drawing from established codes of conduct and ethical standards in other content-creation industries. It delves into the current state of ethical awareness on Large Language Models (LLMs). By dissecting the mechanism of content generation by LLMs, four key areas (upstream/downstream and at user prompt/answer), where safeguards could be effectively applied, are identified. A comparative analysis of these four areas follows and includes an evaluation of the existing ethical safeguards in terms of cost, effectiveness, and alignment with established industry practices. The paper's key argument is that existing IT-related ethical codes, while adequate for traditional IT engineering, are inadequate for the challenges posed by LLM-based content generation. Drawing from established practices within journalism, we propose potential standards for businesses involved in distributing and selling LLM-generated content. Finally, potential conflicts of interest between dataset curation at upstream and ethical benchmarking downstream are highlighted to underscore the need for a broader evaluation beyond mere output. This study prompts a nuanced conversation around ethical implications in this rapidly evolving field of content generation.

## I. INTRODUCTION

ARTIFICIAL intelligence (AI) technology is advancing rapidly, while simultaneously becoming widely accessible to broader society. The speed of this progression reveals oversights regarding ethical awareness in the standards used in Large Language Models (LLM)-based products. These include (i) a lack of content or source attribution, and (ii) a lack of transparency in what was used to train the model. With governments considering regulatory measures for LLM-generated content, LLM-based service providers could draw lessons from other content-producing industries to self-regulate. In addition, in the existing debate around LLM-based AI [1], the implementation of safeguards has primarily focused on output filtering, which overlooks safeguards that could be applied to check the quality of the data used to train the LLMs. This paper will address pedagogy and the adoption of ethics in computer science. It comprises an exploration of possible points of application of standards at both input and output stages of LLMs. These stages are referred to elsewhere as upstream and downstream, respectively. Furthermore, the paper highlights best practices in journalism as one profession that has well-established ethics standards, which could be extrapolated to LLM-based services.

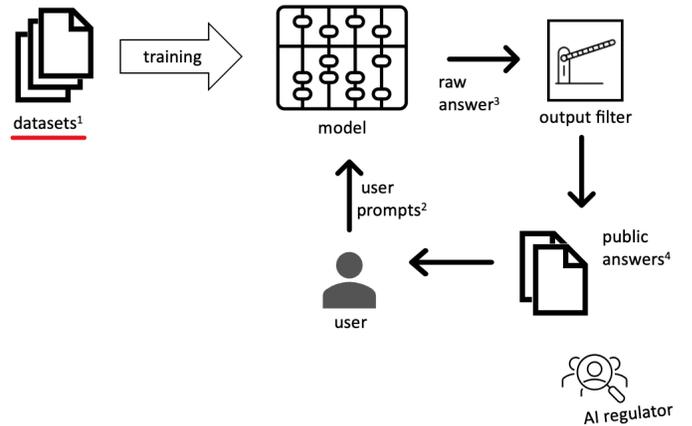

**Fig. 1.** An LLM-based service can be regulated at four points divided in two categories: at the input (1,2), or at the output (3,4).

TABLE I
CODES OF ETHICS BY YEAR

| Field | Document | Year |
|---|---|---|
| Medicine | Hippocratic Oath [e] [11] | (BCE) 400 |
| IT | ACM Code of Ethics [a] [12] | **1992** |
| Civil Eng. | ASCE Code of Ethics [b] [13] | 1914 |
| Biology | Biologists' Code of Ethics [14] | 2004 |
| Chemistry | ACS Code of Ethics [15] | 2012 |
| Physics | AIP Statement of Ethical Principles [16] | 2002 |
| Police | Law Enforcement Code (IACP) [17] | 1957 |
| Firefighters | Firefighters Code of Ethics (IAFC) [18] | 2000 |
| Nursing | Code of Ethics for Nurses (ANA) [c] [19] | 1950 |
| Psychology | APA Ethics Code [d] [20] | 1953 |
| Architects | RIBA Code of Conduct [21] | 2005 |
| Social Work | NASW Code of Ethics [d] [22] | 1960 |

J. Berengueres is with the UAE University, CIT, 17551 Al Ain, AD, UAE (e-mail: jose@ uaeu.ac.ae), M. Sandell is an independent journalist and researcher.

| Academic Publishing | Committee on Publication Ethics [23] | 1997 |
|---|---|---|

a. (latest version in 2018)
b. (latest revision in 2006)
c. (latest revision in 2015)
d. (latest version in 2017)
e. Major revision in 1948. In Nazi Germany, medical students did not take the Hippocratic Oath [24]

### A. IT Ethics awareness post-2005

WASC (Western Association of Schools and Colleges) and ABET (Accreditation Board for Engineering and Technology) both emphasize the importance of incorporating professional responsibility and ethics into the curriculum for engineering, software, and IT programs. ABET has been accrediting programs since 1932, and has always been focused on continuous improvement and evolving standards to meet industry and societal needs. Established later, in 1962, WASC also aims to foster a culture of continuous improvement. One of ABET's criteria for accrediting computing programs is Criterion 3: Student Outcomes, which states that "students must have an understanding of professional, ethical, legal, security, and social issues and responsibilities". This criterion applies to programs in various areas such as electrical engineering, software engineering, and IT [2-4]. ABET and WASC guidelines can influence the content of universities curricula pursuant to their accreditations, thus comprising a major reason why ethics or professional responsibility subjects are included in said programs.

### B. Ethics in IT pre-2005

Interestingly, prior to 2005, we seldom find any reference to ethics in IT curricula. Today, as noted by several authors, most IT programs include some sort of ethics content in their curricula [5-8]. However, compared to other fields, IT has been a relative latecomer to said trend. See Table 1. This is unsurprising. The term 'IT' was only coined after the 1950s (after the invention of the solid-state transistors) and only became mainstream after 1981 with the publication of the PC standard by IBM in the same year.

### C. IT leaders lacking formal ethics training

This broader trend is reflected in the personal academic journey of the first author's professional experience; the first author initially studied as an undergraduate at what is now known as BarcelonaTech from 1994-1999, followed by doctoral studies at TokyoTech from 2005-2007. Throughout his education, several subjects addressed humanistic principles, such as co-existence, but very few provided actionable ethical frameworks, including tools such as introduction to ethical analysis, cost-benefit evaluation, and the like. Fast-forward a decade, he transitioned from student to a lecturer entrusted with teaching an "Ethics for IT" course (ITBP370) from 2014 to 2021. During this period, he employed various educational resources, including Reynolds' textbook on Ethics for IT from 2003 [9], and the interactive Moral Machine website [10], along with case studies and simulated games. This experience led him to two key realizations. First, before his doctoral studies of 2002-2005, he spent several years working in the industry, specifically writing Java code for a German bank, with little formal awareness of ethical considerations. The word "ethics" or "compliance" never surfaced in their teams' discussions; their primary focus was to ensure the code functioned correctly. Second, we postulate that this lack of formal ethics training likely extends to many IT leaders and workers who graduated before 2005.

### D. Awareness as a prerequisite

It is important to note that ethics training does not guarantee ethical behavior. Rather, it serves as a prerequisite for ethical performance [9]. Ethical codes are not a recent development either. Table 1 provides a comprehensive historical timeline. For example, the earliest known code specific to a professional trade can be traced back to Greece in the 5th century B.C. The first recorded ethical code tailored to Computer Engineering was established much later, in 1992. For context, Facebook launched in 2004 and became publicly available outside university campuses in 2006. Meanwhile, the EU General Data Protection Regulation (GDPR) was not enacted until a decade later, in 2016 [25]. The first GDPR-sanctioned fines to Facebook were issued in 2022 for violations in 2018.

## I. CURRENT STATE OF STANDARDS AND ETHICS IN LLM

### A. The data equivalence to the model

Fig. 1 illustrates a simplified information flow in an LLM-based chat service. The model's weights are symbolized by an abacus. Given a specific set of documents for training, most LLMs, such as Facebook's LLaMA [26], and others [27], produce similar "weights" that respond in comparable ways to identical prompts. These weights can be considered a knowledge representation of the underlying training data mediated by user prompts. See Data-Information-Knowledge-Wisdom model in our previous work [28]. From a user's perspective, the model appears to display creativity and "sparks" of abstract reasoning [29]. However, for an informed observer, [30] LLMs are merely a predictor, trained through reinforcement learning to cater to human preferences [31]. This distinction is evident in the HuggingFace LLM leaderboard, a popular platform for comparing LLMs. The leaderboard [27] reveals that the data used for training has a more significant impact on the model's performance than the size of the model (measured in billions of weights) [32]. In essence, the data quality has a more profound influence than the algorithm in the performance. This principle is often referred to in data science as "garbage in, garbage out".

### B. The fine-tunning problem

To increase their practical utility and to align their behavior more closely with human expectations, LLMs are fine-tuned post initial training. This process is a computationally smaller [30], more focused training regime where the model is further refined, usually using a carefully curated dataset comprised of human feedback on the answers received to prompts [33]. Ideally, the model learns to follow instructions more accurately and reduce the likelihood of providing answers that receive poor evaluation by the users (see thumbs up thumbs down image in Table 2). This fine-tuning has proven crucial to commercialize LLM-based services. While the raw models have an impressive ability to understand and generate human-like text, their behavior can sometimes diverge from political correctness. For this purpose, companies such as OpenAI use a method called Reinforcement Learning from Human Feedback (RLHF) to "tune" the raw LLM output [34].

*1) Benefits*

At the heart this fine-tuning is the principle of shaping the model's output to better fit specific, desired characteristics. For instance, model developers might wish to ensure that the AI does not propagate harmful misinformation or express biased views. To achieve this, they might use a dataset specifically designed to promote responsible behavior in AI systems during the fine-tuning process. These datasets could contain examples of appropriate responses to a wide range of prompts, potentially mitigating the risk of harmful outputs. Various studies have indicated that this approach can help in curbing the undesirable outputs of the raw LLM, especially when combined with a comprehensive evaluation mechanism.

*2) Risks*

However, the process is not without challenges. Prompt injection and other techniques can still be effective, even on a fine-tuned model (See "Do Anything Now" hack). In addition, fine-tuning is ideally tailored to the specific contexts and use-cases, therefore, it can also be used to "game" any ethical benchmark. See TruthFulQA in next section. The fine-tuning relies on another dataset, if not disclosed it is naturally hard to oversee any ethics. Note that the finetuning process can itself introduce its own form of bias. Striking the right balance between allowing an LLM to generate diverse outputs and ensuring it adheres to ethical and legal norms is therefore a complex task. (pers. comm., Emil Ahlbäck 2023).

*C. Current standards to compare models*

In the said HuggingFace leaderboard, the models are ranked by a weighted average of four benchmarks. According to HuggingFace, these are:

1) **AI2 Reasoning** Challenge (25-shot)
   This benchmark consists of a set of elementary-level science questions [35].
2) **HellaSwag** (10-shot): Comprises a test of commonsense inference, which is easy for humans (~95%) but poses a challenge for state-of-the-art models [36].
3) **MMLU (5-shot)**
   This measure tests a text model's multitask accuracy across 57 tasks, including elementary mathematics, US history, computer science, law, and more [37]
4) **TruthfulQA (0-shot)**
   This benchmark evaluates the truthfulness of a language model's generated answers to questions [38].

As these, and other ethic-focused benchmarks [39-41] evolve into standards, their influence is anticipated to grow. However, note how in HuggingFace only one of the four benchmarks addresses ethics. In particular, the narrow issue of a model's propensity to regenerate conspiracy theories that were present in the training dataset. Also of note is the lack of concern regarding the provenance of the training datasets [31].

However, what is accepted in IT could be considered a grave violation in other fields such as journalism. As LLM-based services increasingly overlap with journalism, should they not adopt their standards too? What are the potential risks of the lack of integration of ethics in these leaderboards?

*D. Where can we do better?*

Table 2 outlines the possible approaches for self-regulation, which can be implemented across four touchpoints in an LLM-based service. Table 3 compares each of the four touchpoints listed earlier in terms of effectiveness, cost, and risk of misalignment. First, the most effective solution, as proposed by the CEO of Stability [42], advocates addressing potential 'AI misalignment' at its root cause – the training datasets as these are a documented source of bias and other harms. Follows an excerpt from a paper lead by researchers affiliated with Google owned DeepMind… "For example, a dataset of Reddit user comments (used to train an LLM) was found to encode discriminatory views based on gender, religion and race [43]. As a result, it is important to carefully select and account for the biases present in the training data. However, ML training datasets are often collected with **little** curation or supervision and without factoring in perspectives from communities who may be underrepresented [44]…" [45].

However, so far, upstream datasets are not a focus of attention [46]. Perhaps because this choice demands a commitment to transparency. It is this step precisely that the industry leader, OpenAI, has been reluctant to take. Instead, they have chosen to not disclose the specific contents of their training datasets so far. Table 4 compares three leading LLM-based services in terms of ethical safeguards they currently implement.

*E. Hiding problems*

In general, engineering practice discourages the avoidance of addressing root causes of problems [43-45]; a premise that holds true in software engineering too. For example, several views from thought leaders on the subject include:

1) Robert C. Martin, also known as "Uncle Bob", discusses the concept of writing clean code in his 2008 series and book. One characteristic of clean code is its readability and comprehensibility; hence problems should be promptly addressed rather than concealed or disregarded [46].
2) In his 1999 book, Refactoring: Improving the Design of Existing Code, Martin Fowler contends that refactoring is fundamentally the process of identifying and tackling the root causes of code design problems [47].
3) Kent Beck, in his work Test-Driven Development, proposes a software development methodology that stresses writing tests prior to implementation code. This process ensures that problems are swiftly identified and addressed, rather than hidden [48].
4) In 2004, Beck's comprehensive guide to software construction included a discussion on the importance of debugging and testing code. He emphasized that code issues should be detected and rectified, not ignored or concealed [49].

These insights – from software engineering experts – underscore a universally accepted principle in engineering: Ignoring, concealing, or failing to address the root causes is seen as detrimental practice, which can lead to poor-quality code, technical debt, long-term maintenance challenges, and perhaps more concerningly to a moral slippery slope.

*F. Upstream vs downstream safeguards*

Other arguments in favor of ethical controls **upstream** or at input, instead of downstream-only are that:
1) Humans who build the data in the first place deserve fair compensation,
2) Controlling things at source follows the design paradigm of prevention, rather than remedial action, and,
3) It also avoids teaching AIs to "game" the system [50-53];

The main argument in favor of **downstream-only** safeguards is one of cost. While evaluating output is straightforward with benchmarks, evaluating the input (training data) for misinformation, conspiracy theories, etc. is more expensive. Once a "poisoned" document is discovered in the training set, engineers do not know how to "remove" its effects on the model weights. The only way forward at this stage is to retrain the model from zero, which takes various days and several millions of dollars in computational costs [30].

TABLE II
TOUCHPOINTS WHERE LLMS CAN BE REGULATED

|  | At input (upstream) | At output (downstream) |
|---|---|---|
| At user level | Prompt censoring | User reports * |
| At model level | Dataset curation | Filtering answers |

*image source: chat.openai.com

TABLE III
COST-BENEFIT ANALYSIS

| | Safeguard touchpoint | Risk addressed | Effectiveness | Cost |
|---|---|---|---|---|
| Upstream | Prompt censoring | Prompt injection, jail break | <100% (See "Do Anything Now" hack) | ✅ Low |
| Upstream | Dataset curation | Regenerating unethical patterns in **training** data | Common sense [42] | ⚠️ High |
| Downstream | Output filter | **Censor** problematic outputs | ⚠️ Hides problem [b] | ✅ Low |
| Downstream | User reports [a] | **Detect** problems that passed the previous filter | Feedback used primarily for RL (not ethics) [30] | ✅ Low |

a. image source: chat.openai.com
b. hides the problem recognized as poor practice in engineering [80-90]

TABLE IV
SAFEGUARD MEASURES BY VENDOR

| Ethics check | OpenAI | Midjourney | Stability | Potential issue |
|---|---|---|---|---|
| Prompt censoring | ✅ | ✅ | ✅ | Hides the problem |
| No use blanket disclaimers | ❌ | ✅ | ✅ | Slippery slope |
| Train data disclosed? | ❌ | ❌ | ✅ | Potential use of content without crediting sources [a] |

a. Stable Diffusion is "trained" on a 100TB dataset that contains 2bn images, including copyrighted photos.

*G. Symptoms of slippery slope*

*1) Use of blanket disclaimers*

See Fig. 2 for an example of the slippery slope – a well-known concept. The term refers to a type of argument where a specific decision or action leads to a series of events that result in an undesirable outcome, the said "slippery" descent into unethical or immoral behavior. This slippery slope phenomena is more prevalent than reported in the media (see for example, [54-63]). In Fig. 2, the statement does not seem in compliance with a few items in the ACM code of ethics, see [12].

*2) Shifting blame to user*

This blanket statement uses a technique called blame shifting [64-67] where users are coopted into shouldering responsibility of misinformation and/or unethical responses. This practice would be similar to a company asking the user to help improve the products but without financial compensation. A further point is that these LLM products have been rolled out to the public, skipping the common practices of beta testing and slow progressive rollouts.

*3) What if the AI was a journalist?*

If the chatbot in Fig. 2 was subjected to the same standards of journalism, such blanket statements would not be allowed. As one user exclaimed, this disclaimer is "equivalent to the New York Times posting a statement on the front page that its content may be wrong, but the NYT isn't responsible. Sorry". With this comparison front of mind, we now turn to exploring how the news industry deals with setting standards for content.

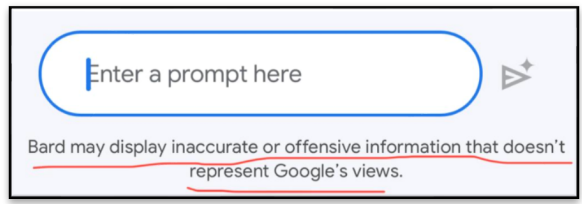

**Fig. 2.** A chatbot blanket disclaimer example. Source: Google.

### III. What AI-owners Can Learn from Journalism

The process of advancing AI technology and making AI-generated content attractive to and available to the general public is comparable to that of the printing press and its pivotal influence on newspapers and journalism. Like AI-owners today, publishers at the time did not start off with a full-fledged set of standards and ethics to guide them. Some newspaper owners pushed for sensational content to attract paying readers, resulting in the creation of the derogatory term "yellow journalism" to describe inaccurate or misleading content [72]. To avoid content that would cause harm, journalists and newspaper owners agreed to set standards for content creation that benefited the consumer of the content and built their own reputations as reliable [73]. Here are some common tenants in these ethical codes that hold relevance for owners and users of AI-generated content.

#### A. Accuracy

Editors have argued that the single most important thing in journalism is accuracy. The editor-in-chief emeritus at Bloomberg News, Matthew Winkler, interviewed and hired hundreds of reporters across his career. He usually asked one question that would result in an automatic rejection if the candidate got it wrong. The question was "What is the most important thing in journalism?" The right answer was "accuracy", which one author of this paper got right before her 18-year career at the organization. Accuracy and factual reporting are the backbone for building and maintaining trust with the content consumer. Accuracy is ensured through mechanisms including:

*1) Editors*

Editors serve as gatekeepers for content. They are the first line filter after the journalist. Editors will monitor for grammar and factual accuracy. They will employ a series of flags to check for accidental inaccuracy. Some editorial tasks are programmed right into the desktop tools, while most traditionally remain with human editors.

*2) Fact-checking*

Publications can have separate departments that check the accuracy of statements through researching databases and calling sources to confirm information. Indeed, some publications have fact-checking reporters who fact-check other news and publish their findings.

*3) Sourcing*

Publications have specific rules for sharing their sources. Bloomberg, for example, requires citing a source before publishing. Exceptions are considered when revealing a source may cause them harm. This decision often requires a rigorous review process and approval from an executive editor [65].

*4) Ombudsman*

Publications employ ombudspersons, who handle complaints from the public. They also have a broad mandate to maintain ethical standards within an organization.

*5) Fireable offenses*

Journalists who do not adhere to these standards are dismissed. Jayson Blair, formerly of the New York Times, was dismissed for fabricating content, events and sources. In an attempt to restore its reputation, the NYT revealed details of what Blair did to deceive the readers and how NYT will endeavor to ensure similar situations do not arise [70].

*6) Transparency*

Transparency is about showing your work and your sources so the consumers of the content can make their own decisions about the information. Transparency gives the consumer power to check accuracy themselves across various media. In research, transparency is usually revealed in footnotes and citation practices. In journalism, it is part of the text, and can be revealed as follows:

*7) Citing sources*

Citing sources means revealing the name and location of the source of the material and information used. If the source content is digital then it necessitates a link directly to the source.

*8) Revealing omission*

When information that may be pertinent cannot be secured, journalistic standards require revealing details about why the information could not be secured. A typical example is when the reporter writes a person "did not respond to requests for an interview".

*9) Transparency of ownership*

If the publication is owned by a party that may have an interest in the topic, or may gain or lose something because of the news, this fact must be revealed.

*10) Conflict of interest*

All potential conflicts of interest are expected to be revealed or avoided. A reporter should not interview their own relative for a story or use a good friend as a source of information.

#### B. Do-no-harm principle

Similar to the Hippocratic Oath, journalists follow a principle of conducting interviews and research so as not to use information gathered in a manner that would harm the people involved. Academic research has similar guidelines for its information gathering processes.

*1) Use of adjectives and adverbs*

Journalists are trained to avoid using adjectives and adverbs that are not clearly backed up with fact to avoid unintended bias. Calling something "tall" is relative, and may be misconstrued from its intended meaning. Saying something is 165 centimeters explains exactly what it is to a person who is 200 cm as well as one who is 150 cm. It is "tall" to one but actually "short" to the other. So, all adjectives and adverbs are potential flags for misleading content.

*2) Clarity between opinion and fact*

The news profession often uses labels for clarity. An opinion piece or a column is clearly labelled Opinion and not News. Labels can be used to show accuracy and provide transparency.

*3) Use of personal information*

When reporting court cases, for example, journalists also use labels to define a defendant as alleged so to ensure readers do not inadvertently believe they are guilty. Also, when writing about a person, the reporter endeavors to give that person time to comment or respond before publication.

*C. Laws and accountability*

Like other industries, there are laws that govern content creation by journalists. These often revolve around libel. Publishers can take out libel insurance to cover their exposure to risk. In AI-generated content, it remains unclear where accountability lies. Will libel for AI-generated content be accountable to the maker of the model or the user of the model?

## I. DISCUSSION

By comparing the ethical standards awareness in section 1 with the already established codes of conduct in journalism addressed in section III, AI-owners can begin to consider what elements might apply to AI-generated content. Section II explores where in the AI-generation process the standards from journalism could be applied. To be sure, this paper is not attempting to decide what is and is not ethical for LLM-generated content. Rather it is suggesting that adopting fundamental best-practices in transparency and accuracy will allow for ethical assessment and ethical-based decisions in content creation -- for both provider and user.

*A. Case where journalism standards would have been effective*

An example worth considering as a discussion prompt is the lawyer who recently sued an airline on behalf of the client [71]. The lawyer submitted a brief that included a number of relevant court decisions. However, it was later revealed that no one could actually find the decisions cited in the brief. The lawyer had used OpenAI's ChatGPT to do his research. Ironically, the lawyer even asked ChatGPT to verify the cases were real. The program confirmed they were real. But they were not; a fact confirmed when the judge went looking for them and found nothing. They were fabricated by AI. Had this been content created in journalism, many of the tools used to guarantee accuracy would have caught the erroneous information before publication. For AI-owners, the tools could be applied to the point of entry of data into the model, the model itself or to post-production filters. AI-owners could also create an ombudsperson system for fielding reports of erroneous or harmful content from end-users. This example provides just one instance that can prompt more nuanced ethics-based discussions. AI-owners can look to other industries that deal with content for inspiration for others [72,73]. We encourage those in other industries such as the real estate or legal professions, where content is created for contracts, to join us in exploring what can be extrapolated from their standards to help benefit this new AI industry. In fact, any industry that could incur risk with their actions can hold ethical inspiration.

## II. CONCLUSION

LLM-bases services generate content. Currently, this content is facing a backlash when it is incorrect, misleading, or potentially dangerous to society. AI-owners can remove some of the concern by creating a code of conduct or standards for their products. None of these issues related to the control or misuse of information are new. They have been faced before in other industries.

This paper has discussed the current state of ethics in computer science, where standards could be applied in the current state of AI-generated content, and how another industry – journalism and publishing – developed effective codes of conduct and standards for addressing content-specific issues of accuracy, transparency, and conflict of interest.

The paper shows there has been a lack of training in ethics in higher education among IT leaders who graduated before 2005, and discussed the influence of WASC and ABET in IT program curricula.

It has discussed a number of touch points in the LLM processes that can be exposed to safeguards and control measures to improve accuracy and transparency, both at the entry of the data – upstream – and after the generation of content – downstream.

It has been noted that there has been a tendency to put efforts into the downstream controls only. This paper raises the question: Why? And why not apply efforts in upstream too? Is there a conflict of interest here for the AI-owners if upstream checks are more costly? As the leading academic institutions in AI also depended on funding from big tech, is this conflict of interest limited to LLM service providers only?

Finally, we listed some principles from journalism for insight into which standards have historically been effective in dealing with the same issues LLM-based generated content is facing today.

Further discussion and research are needed into the usefulness and application of established standards in other content-creating industries. The key motivation is that AI-owners minimize harm and promote accuracy and transparency, thus inserting some of the fundamental standards and behaviors needed to advance ethical AI.

## III. FINAL REMARKS

Journalism's codes of ethical conduct have developed, and been refined, over several decades. These processes of refinement have resulted in a system of content generation with high level of public trust and reliability. Yet, such trust did not come overnight, it took time. The utility of using journalism as a comparison in this ethics-based discussion manifests in two points relating to speed and trust. We know that the speed at which AI based content is developing is exceeding expectations; with experts in the computer science field calling for a pause its development for six months to allow government regulators to 'catch up' [74]. This call is indicative of the widespread concern for its rapid, and somewhat ethically unfettered development. Second, and corollary to the first, is trust. As regulators scramble to address the daily array of concerns regarding misinformation and the veracity of AI-generated content, trust in the information produced fluctuates. In the context of the computer science, such trust issues are compounded by the fact the discipline does not have a robust foundational base in ethics education to springboard from.

ACKNOWLEDGEMENT

Danielle Drozdzewski for editorial assistance.

**J. Berengueres** (Member, IEEE), photograph and biography not available at the time of publication.

**M. Sandell**, photograph and biography not available at the time of publication.